\documentclass[preprint,12pt]{aastex}
\usepackage{epsfig}
\usepackage{emulateapj5}
\usepackage{apjfonts}

\newcommand{\chandra}{{\it Chandra}}

\newcommand{\rosat}{{\it ROSAT}}

\newcommand{\lum}{\thinspace\hbox{$\hbox{erg}\thinspace\hbox{s}^{-1}$}}

\begin{document}

\def\spose#1{\hbox to 0pt{#1\hss}}
\def\laeq{\mathrel{\spose{\lower 3pt\hbox{$\mathchar"218$}}
     \raise 2.0pt\hbox{$\mathchar"13C$}}}
\def\gaeq{\mathrel{\spose{\lower 3pt\hbox{$\mathchar"218$}}
     \raise 2.0pt\hbox{$\mathchar"13E$}}}

\title{Discovery of Radio/X-ray/Optical Resolved Supernova Remnants in
the Center of the Andromeda Galaxy} 

\author{A.K.H.~Kong}

\affil{Harvard-Smithsonian Center for Astrophysics, 60 Garden Street,
Cambridge, MA 02138; akong@cfa.harvard.edu}

\author{L.O.~Sjouwerman}
\affil{National Radio Astronomy Observatory, Socorro, NM 87801}

\author{B.F.~Williams, M.R.~Garcia}
\affil{Harvard-Smithsonian Center for Astrophysics, 60 Garden Street,
Cambridge, MA 02138}

\author{and}

\author{J.R.~Dickel}
\affil{Department of Astronomy, University of Illinois at Urbana-Champaign, Urbana, IL 61801}

\label{firstpage}

\begin{abstract}
We have detected a spatially resolved supernova remnant (SNR) in the
center of the Andromeda Galaxy, in 
radio, X-ray, and optical wavelengths. These observations provide the highest spatial
resolution imaging of a radio/X-ray/optical SNR
in that galaxy to date. The multi-wavelength morphology, radio spectral
index, X-ray colors, and narrow-band optical imaging are consistent
with a shell-type SNR. A second SNR is also seen resolved in both
radio and X-ray. 
By comparing the morphological sturcture of the SNRs in
different wavelengths and with that in our own Galaxy, we can study
the shock morphologies of SNRs in the Andromeda Galaxy.
The proximity of the SNRs to the core 
suggests  high interstellar
medium density in the vicinity of the SNRs in the center of the Andromeda Galaxy.

\end{abstract}


\section{Introduction}

Ground-based and spaced-based observations have allowed many
supernova remnants (SNRs) in our Galaxy to
be resolved, showing various different morphologies in radio, X-ray,
and optical wavelengths (Weiler \& Sramek 1988; van den Bergh 1988;
Rho \& Petre 1998). 
The expanding SNR
shock can heat the surrounding interstellar medium (ISM) to temperatures
up to $10^8$ K and beyond, producing thermal X-ray emission, while
synchrotron radiation seen in the radio band is due to the relativistic
electrons accelerated behind the shock front in the magnetic field
of the SNR. Optical emission from SNRs is from shocked-heated collisionally ionized
species such as [S II], [N II] and [O III], as well as H$\alpha$
recombination emission. Multiwavelength observations of SNRs
therefore provide important information on the structure of the ISM and chemical
composition in the host galaxy, the evolutionary stage of the SNR, the
character of the
supernova ejecta and the nature of the progenitor and its associated stellar winds.

Most detailed studies of SNR have concentrated on our Galaxy simply because of
the large angular size of the SNRs. However,
studies of Galactic SNRs can be limited by the lack of reliable
distance estimates, high interstellar absorption at optical and X-ray
wavelengths, and the high level of confusion in many Galactic
fields (Magnier et al. 1995). These limitations are overcome by
studies of SNRs in nearby extragalactic 
systems such as the Large and Small Magellanic Clouds (e.g., Williams et
al. 1999), and the
Andromeda Galaxy (M31). At a distance of 780 kpc (Stanek \& Garnavich 1998), M31
is the closest spiral galaxy that
shares similar morphology, metallicity and size with the Milky Way; it
is the best candidate for a comparative study of SNRs. 

Studies of SNRs
in M31 are mainly done in the optical (e.g., d'Odorico,
Dopita, \& Benvenuti 1980; Blair, Kirshner, \& Chevalier 1981, 1982;
Braun \& Walterbos 1993; Magnier et al. 1995). About 200 SNR candidates
have been identified by these surveys. In a recent
extensive \rosat\ Position Sensitive Proportional Counter (PSPC) survey (Supper et al. 2001) of
M31, 16 X-ray SNRs were identified within a total area of $\sim 10.7$
deg$^2$ by cross-correlating with these
optical catalogs. Their X-ray luminosities (0.1--2.4 keV) range from $10^{36}$ to $10^{37}$ \lum.
More recently, Kong et al. (2002a) discovered a spatially
resolved X-ray SNR in M31 with \chandra. In the radio, Sjouwerman \&
Dickel (2001) found 3 SNRs near the center of M31. In this {\it Letter}, 
we report
the discovery of the first radio/X-ray/optical resolved SNRs in the
center of M31.

\section{Observations and Results}
\subsection{Radio Data}

At the distance of M31, the typical diameter of SNRs is between $\sim 1$
arcsec and 20 arcsec so that some of the 
SNRs are large enough to be studied with high-resolution
instruments in sufficient detail (Sjouwerman \&
Dickel 2001; Kong et al. 2002a).  In particular, sensitive
interferometric radio images have sufficient resolution to resolve the
SNRs without suffering from the large scale diffuse background in the
center of M31 (Hjellming \& Smarr 1982; Crane, Dickel, \& Cowan 1992;
Crane et al. 1993). 
Very sensitive 8.4 GHz (3.6 cm) radio observations of the
central 500 arcsec (1.9 kpc) of M31 were obtained by combining in
total 120 hours of archival data taken with the Very Large Array (VLA)
in its largest array configurations (A and B) between 1990 and
1996. The resulting radio image has an angular resolution of 0.24
arcsec and is sensitive to angular sizes of up to 20 arcsec, and
has an rms noise level of $\sim$ 4 $\mu$Jy/beam. After discovering
three faint radio SNRs in the 8.4 GHz archival data, we also observed
the remnants for 24 hours with the VLA at 4.9 GHz (6 cm) in July
2002. The recent 4.9 GHz radio image (Fig. 1) is sensitive to angular
structures from 1.2 to 36 arcsec and has an rms of $\sim$ 5
$\mu$Jy/beam. Both radio images clearly show resolved shell-like objects
that are brighter at 4.9 GHz than at 8.4 GHz (Sjouwerman \&
Dickel 2001). The largest SNR is located at 70
arcsec east-south-east of the nucleus of M31 at a position of
R.A.=00$^{\rm h}$42$^{\rm m}$50$^{\rm s}$.41,
Dec.=+41$^{\circ}$15$^{\prime}$56$^{\prime\prime}$.4 (J2000), labeled
object 1 in Fig. 1. 
The integrated flux density of the complex radio morphology is
measured to be 1.13
mJy at 4.9 GHz. Its size of $11.6\times8.4$ arcsec ($44\times32$ pc)
is consistent
with a previous $10.9\times7.8$ arcsec low-resolution (5 arcsec) VLA detection
at 1.4 GHz designated Braun-101 (Braun 1990). With the notion that the flux density
measured at 1.4 GHz might be confused with the extended background
(Hjellming \& Smarr 1982; Braun 1990),
we derive a spectral index between 1.4 and 4.9 GHz of $\alpha$ = $-$0.77 or
flatter, where $S_\nu \propto \nu^\alpha$. The radio morphology and radio 
spectral index are consistent with a shell-type SNR (Dickel \& Milne 1998).
It is worth noting that our radio observations also discover two more new SNRs, designated
Braun-80 and Braun-95 (Braun 1990), near the nucleus of
M31 (Sjouwerman \& Dickel 2001). 

\subsection{X-ray Data}

Following the discovery of the radio SNRs, we compare their radio positions
with the M31 X-ray source positions as detected by the \chandra\ X-ray
Observatory (Kong et al. 2002b; Kaaret 2002) and indeed the largest
radio SNR, Braun-101, is spatially coincident 
with the X-ray source CXOM31 J004250.5+411556 (Kong et al. 2002b). We further examine a
40 ksec (37.7 ksec after rejecting high background period) \chandra\
X-ray image taken with the back-illuminated chip (S3) 
of the Advanced CCD Imaging Spectrometer (ACIS). 
The SNR is clearly detected in the 0.3--7 keV image and it is
resolved as a ring-like object with a diameter of about
5 arcsec (Fig. 1). The SNR has about 55 background-subtracted
photons (0.3--7 keV) in a $3''$ radius extraction region. All of the emission from
the SNR is 
confined to energies
below $\sim 1$ keV, indicating that it is a soft X-ray
source (e.g., Di\,Stefano \& Kong 2003). The shell-like X-ray morphology, soft
X-ray nature, and good positional
coincidence with the radio SNR makes the X-ray source as a secure X-ray counterpart
of the radio identified SNR.  We extracted the energy spectrum from a
$3''$ radius and background from an annulus centered on the source. In
order to allow $\chi^2$ statistics to be used, the spectrum was grouped into
at least 10 counts per spectral bin. We fitted the data with a
Raymond-Smith model and kept the $N_H$ fixed at $10^{21}$
cm$^{-2}$. The best fit
temperature is $0.25^{+0.09}_{-0.06}$ keV (90\% confidence) with a
value of reduced chi-squared statistic $\chi^2_{\nu}=1.05$ for 5 degrees of freedom (d.o.f.).
The 0.3--7 keV luminosity of the best fitting spectrum is
$(4.4^{+1.3}_{-2.2})\times10^{35}$\lum. We also 
re-examined the 50 ksec \chandra\ High Resolution Camera (HRC) observation
(Kaaret 2002) and Braun-101 is indeed visible as a faint extended source though
it was not in the published HRC source list. Assuming the above
spectral model, the corresponding 0.3--7 keV luminosity of the SNR is
$(7.1\pm2.5)\times10^{35}$\lum.

The other radio SNR, Braun-95, is also detected in the
\chandra\ observation as an extended (with a diameter of $\sim 10$ arcsec) object (marked as 2 in
Fig. 1). Although it shows a clear shell morphology in the
radio image, the X-ray emission is very patchy. We extracted the energy spectrum with
a $4.5''$ radius circle centered on the source and background from a
source-free region. We kept the $N_H$ fixed at $10^{21}$ cm$^{-2}$
and fitted the spectrum with Raymond-Smith model. The fit is rather
poor ($\chi^2_{\nu}=1.55$ for 10 d.o.f.). We then added an
additional power-law component and the fit was improved 
($\chi^2_{\nu}=1.2$ for 8 d.o.f.). The best fit temperature is
$0.26^{+0.13}_{-0.07}$ keV, while the photon index is
$2.16^{+1.39}_{-2.16}$; the 0.3--7 keV luminosity is
$(8^{+7.0}_{-2.2})\times10^{35}$\lum. 

\subsection{Optical Data}

With the detection of radio and X-ray emissions from Braun-101, we
searched the {\it Hubble Space Telescope} ({\it HST}) archive for any narrow-band
exposures near that position. We found images taken with the Wide Field
Planetary Camera 2  (WFPC2) in H$\alpha$ (F656N) filter. We downloaded
the pipeline-calibrated images, registered them, and combined them
with cosmic-ray rejection. The resulting continuum (F547M filter) subtracted image of
the region around the SNR is shown in Figure 2.
A resolved
shell-like  object with 3 arcsec 
diameter is seen at the radio and X-ray position in the
H$\alpha$ image, indicating an
optical counterpart
of the SNR. The total H$\alpha$ flux is $(7.3\pm0.5)\times10^{-16}$ erg
cm$^{-2}$ s$^{-1} \mathring{\rm A}^{-1}$, while the peak H$\alpha$ surface
brightness of the SNR is $(3.3\pm0.4)\times10^{-16}$ erg cm$^{-2}$ s$^{-1}$
arcsec$^{-2} \mathring{\rm A}^{-1}$ in the continuum-subtracted H$\alpha$ image. 
Re-examination of ground-based narrow-band optical images from the Kitt
Peak National Observatory 0.6-m Burrell Schmidt telescope also
reveals Braun-101 (but unresolved) with [S II]/H$\alpha\sim
0.65$. This high ratio makes it unlikely
that the H$\alpha$ emission is from a HII region near the SNR
(Levenson et al. 1995), and
confirms the radio/X-ray identification as a SNR.
Unfortunately, radio/X-ray SNR, Braun-95, is just outside the field of view of the
{\it HST} observations.

\section{Discussion}

Using VLA, \chandra, and {\it HST} data, we found one
radio/X-ray/optical resolved SNR (Braun-101) and one radio/X-ray
resolved SNR (Braun-95) near
the center of M31. 
The radio emission of Braun-101 traces the outer shell of the remnant; it
is an asymmetric shell with the northeastern side about three times brighter
(67 versus 23 $\mu$Jy/beam) than the southwestern side.
The X-ray
remnant  with a diameter of
about 5 arcsec (20 pc) is considerably smaller than the radio remnant and the
X-ray outline lies predominantly inside the brighter part of the
radio remnant. Both wavelengths show a shell configuration but the X-ray remnant is
elongated toward the north, while the southern part of the X-ray
remnant near the center of the radio remnant is slightly brighter in
X-ray. 
It could be that the situation is similar to that seen for SNR
1E\,0102.2-7219 in the Small Magellanic Cloud (Amy \& Ball 1999; Gaetz
et al. 2000). 
The X-ray emission is
brightest in a ring primarily 
inside the radio shell but very faint X-rays are seen extending out to the shock 
just outside the radio shell.  The bright inner X-ray ring is interpreted as 
coming from heating and ionization from the reverse shock moving into a strong 
gradient in pre-explosion ejecta of the progenitor star.  
A longer X-ray exposure might reveal the expected faint outer X-ray
emission. The larger radio shell is perhaps the forefront
of the shock in a relatively low ISM density toward southwest, where the
greatest magnetic field compression and particle acceleration
occurs. 

The optical emission of Braun-101 is the smallest
(3 arcsec; 11 pc) compared to other wavelengths and it is encompassed by
the X-ray emission (Fig. 2). The H$\alpha$ shell roughly traces the inner
X-ray ring of the remnant and shows a non-uniform
distribution of emission over the shell, with slight enhancement toward
the southwest. It also shows a crescent opening and possibly a
double-ring structure in the northern region of the optical
remnant. In the northeast, a local enhancement of the X-ray emission
corresponds to a local optical enhancement. 
The optical emission could
be produced by shocks driven into 
cooler and denser material by the high pressure downstream of the
reverse shock. It is worth noting that such large discrepancies
in SNR size at different wavelengths are not typical (Winkler \& Long 1997). In this case,
we may be seeing only the brightest optical knots and not the whole
optical shell.

Braun-101 is only 0.5 arcsec from an optically identified planetary
nebula candidate
(Kong et al. 2002b; Ciardullo et al. 1989). Our multiwavelength observations provide strong evidence that
the source is actually a SNR. As planetary nebulae are usually
identified by their strong [O III] emission (Ciardullo et al. 1989), they could resemble the
[O III] emission from some SNRs. However, the large angular size
(from 10 pc in the optical to 40 pc in the radio) and high X-ray
luminosity ($4.4\times10^{35}$ erg s$^{-1}$) rule out the possibility
of a planetary nebula for which the typical diameter and X-ray
luminosity is $< 1$ pc ($< 0.25$ arcsec) and $10^{30}$ erg s$^{-1}$, respectively. 

Like Braun-101, Braun-95 is clearly a shell-like radio remnant (see
Fig. 1) and its morphology is also very similar to Braun-101. 
The X-ray remnant of Braun-95, however, is patchy and is mainly located in the
northeastern side of the radio remnant, where the radio emission is
brightest. Faint X-ray emission is also seen near the radio shell. The
X-ray spectrum likely consists of two components: a Raymond-Smith model and a
power-law model. The Raymond-Smith temperature is similar to Braun-101
and CXOM31 J004327.7+411829 (Kong et al. 2002a). The extra power-law
component might hint that Braun-95 contains a pulsar and/or a pulsar nebula.

Discovery of SNRs near the center of M31 provides some information
about the star formation history of the region. For instance, type Ia
SNRs suggest an old stellar population, while type Ib/II SNRs are
associated with young population. Unfortunately, we do not have direct
evidence about the type for the two SNRs based on the current
multi-wavelength data. By comparing with Galactic
SNRs, the mismatch in morphology of the two SNRs in different wavelengths might hint
the type. For example, IC 443 (Petre et al. 1988; Kawasaki et
al. 2002) and Sgr A East (Maeda et al. 2002) are type II SNRs that
have different morphology in radio and X-ray, and type
Ia SNR like Tycho and SN 1006 generally 
have a well matched morphology in X-ray and radio. On the other hand, the
X-ray and radio surface brightness of Tycho does not show
correspondence (Hwang \& Gotthelf 1997), and the X-ray emission of
SN 1006 is non-thermal (Long et al. 2003), suggesting a synchrotron
origin of X-rays. In addition, type Ib/II SNRs
such as Cas A (compare Hughes, et al. 2000 and Braun, Gull \& Perley
1987; see also Keohane et al. 1996 and Dickel et al. 1982) and
RCW103 (compare Dickel et al. 1996 and Tuohy \& Garmire 1980) also
have a good match in different 
wavelengths. 
Further multi-wavelength observaitons of these 2 SNRs are
required to identify their type and hence the assoicated stellar
population in the bulge of M31. 

The multiwavelength detection of SNRs in M31 can also provide further
information about the local ISM density around the SNRs. The ISM near M31's SNRs is
thought to be low ($< 0.1$
cm$^{-3}$), which is 1 to 2 orders of magnitude smaller than the
measurements by the H\,I observations (Magnier et al. 1997). Our discovery of X-ray
emission from SNRs near the nucleus of M31 indicates that the ISM density
in the vicinity of the SNRs could be higher.

We have shown that by combining all the high-resolution instruments
across many wavebands, we can study the morphological structure
of SNRs in M31 for the first time. This is a first step toward understanding the evolutionary
states and shock morphologies of SNRs in M31. More specific multiwavelength
observations in the future will
shed more light on the properties of SNRs in M31, allowing more
comprehensive comparative studies with our own Galaxy.

\begin{acknowledgements}
This research was supported by NASA under LTSA grants,
NAG5-10889 and NAG5-10705. A.K.H.K. acknowledges support from the
Croucher Foundation.
\end{acknowledgements}

\begin{figure}
\psfig{file=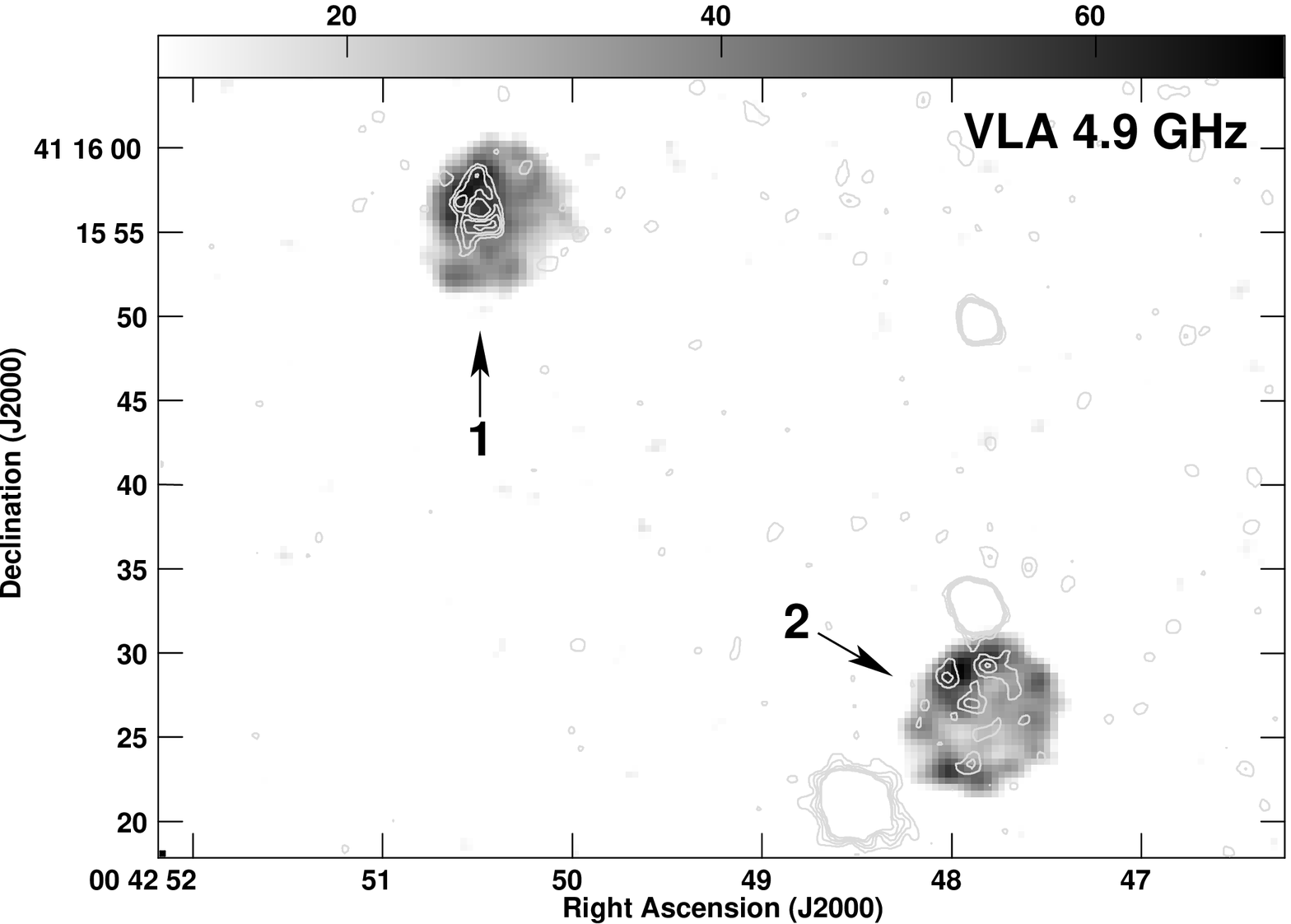,height=10cm}
\psfig{file=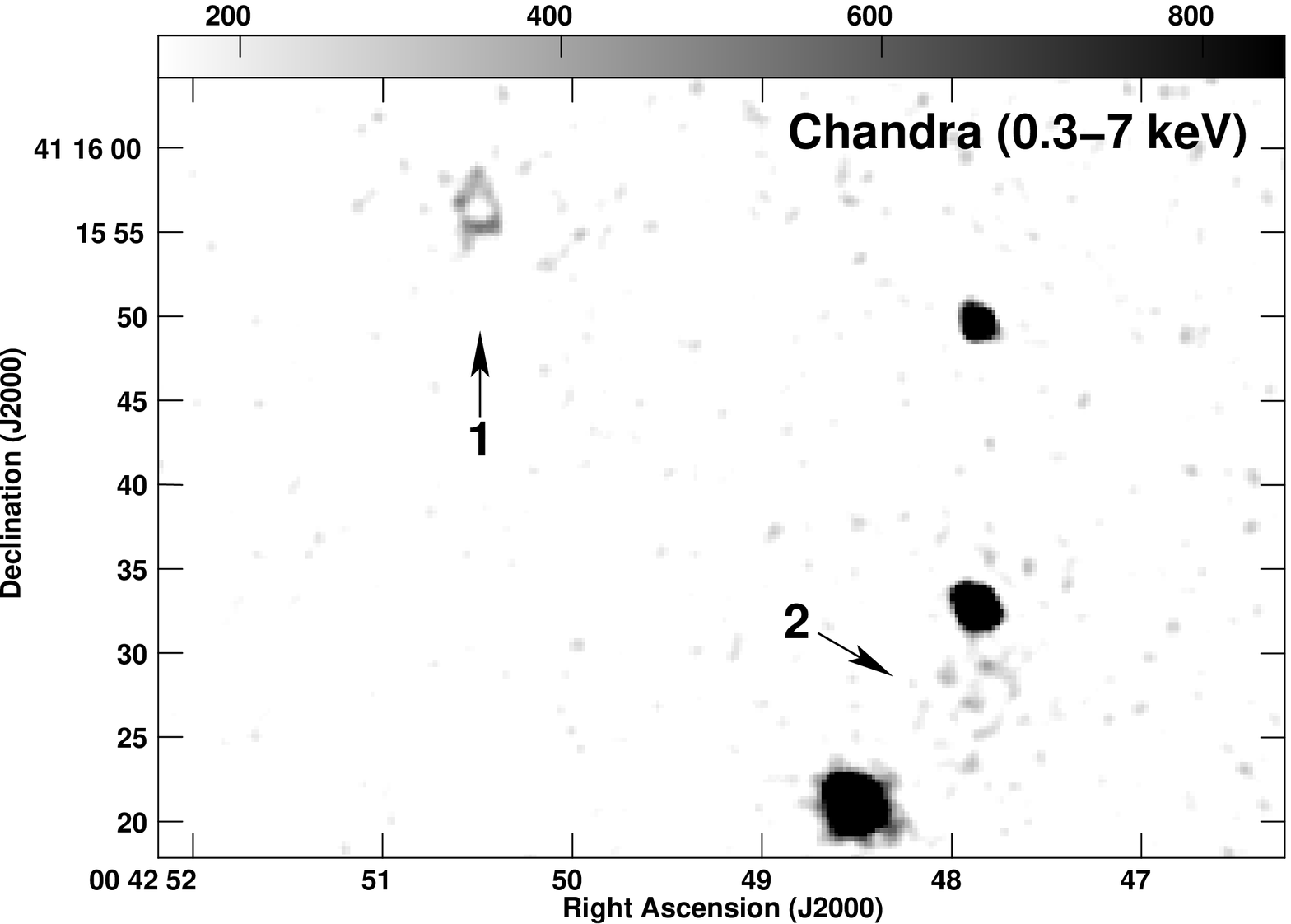,height=10cm}
\caption{Top: VLA 4.9 GHz (1.2 arcsec resolution) radio image (grey
scale from 10 to 70 microJy) of CXOM31
J004250.5+411556 (marked as 1) and nearby region with the \chandra\ X-ray
contours. The contour map was created from a smoothed image with a 0.5
arcsec $\sigma$ Gaussian function. Contours are at 0.6, 0.9, 1.2, and
1.5$\times10^{-5}$ counts sec$^{-1}$ arcmin$^{-2}$. Also
shown in the figure is another radio SNR, Braun-95 (Sjouwerman \&
Dickel 2001; Braun 1990), on the southwestern side
(marked as 2). The radio position of the SNR is R.A.=00$^{\rm h}$42$^{\rm m}$47$^{\rm s}$.82,
Dec.=+41$^{\circ}$15$^{\prime}$25$^{\prime\prime}$.7 (J2000).
To match the radio and X-ray images, we used the
nucleus of M31 (M31*) to cross-register both images. The two arrows
have a length of $5''$. Bottom: \chandra\
0.3- to 7-keV image
of the same
field as the above radio image. The
image has been smoothed with a 0.5 arcsec $\sigma$ Gaussian function. The
scales of both images are the same.}
\end{figure}

\begin{figure}
\psfig{file=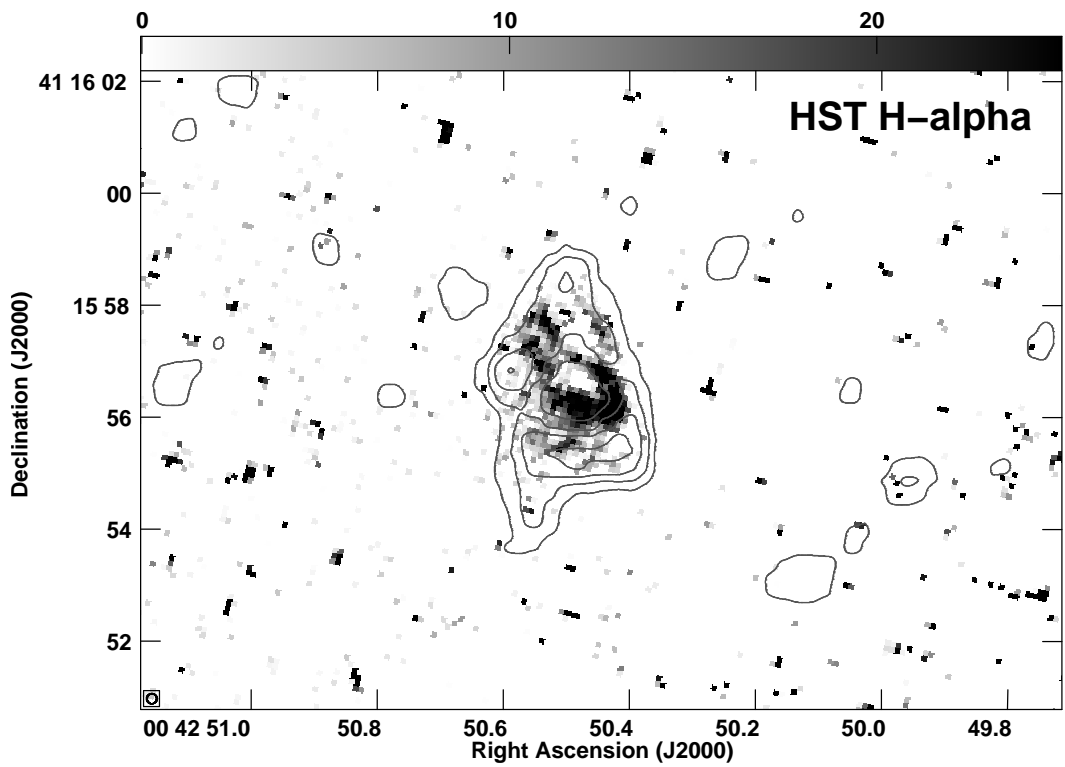,height=10cm}
\caption{{\it HST} continuum-subtracted H$\alpha$ image 
of the blow up region of CXOM31 J004250.5+411556 with the \chandra\ X-ray
contours (same as Fig. 1). We used the X-ray position of M31* (Kong et
al. 2002) and optical position of the double nucleus P2 (Lauer et
al. 1993) to match both images.} 
\end{figure}

\end{document}